\documentclass[12pt]{article}
\usepackage{latexsym}
\usepackage{graphicx}
\title{Solipsistic hidden variables}
\author{Hrvoje Nikoli\'c \\
Theoretical Physics Division, Rudjer Bo\v{s}kovi\'{c} Institute, \\
P.O.B. 180, HR-10002 Zagreb, Croatia \\
{\normalsize e-mail: hrvoje@thphys.irb.hr} \\
\makebox[1in]{} \\
}
\date{\today}
\begin{document}
\maketitle
\begin{abstract}
We argue that it is logically possible to have a sort of both reality and locality
in quantum mechanics. To demonstrate this, 
we construct a new quantitative model of hidden variables (HV's),
dubbed solipsistic HV's, that interpolates between 
the orthodox no-HV interpretation 
and nonlocal Bohmian interpretation. In this model,  
the deterministic point-particle trajectories are associated only with the essential 
degrees of freedom of the observer, and not with the observed objects. 
In contrast with Bohmian HV's, nonlocality in solipsistic HV's can be substantially 
reduced down to microscopic distances inside the observer. Even if such HV's
may look philosophically unappealing to many, the mere fact that they are
logically possible deserves attention. 
\end{abstract}


\section{Introduction}

The no-local-hidden-variable theorems \cite{bell,GHZ,hardy}
for quantum mechanics (QM)
have profound, but not unambiguous \cite{nikmyth}, 
implications on the nature of objective physical reality --
reality supposed to exist even without observations.
Two typical but mutually confronting views inferred from these theorems are (i) that
nature is local but objective reality does not exist \cite{mermcor,zeil,rov2}, or 
(ii) that objective reality exists but is not local \cite{bell,durretal,norsen}.
Moreover, many seem to agree that an intermediate option, 
which would retain both objective reality and locality, is not possible.

With a motivation to reduce the confrontation between these two views,
as well as to demonstrate that an intermediate option is at least not impossible,
in this paper we propose a new ``hybrid'' approach.
In this approach, some elements of each of the two options are combined
into a new interpretation that, to a certain extent, retains both
objective reality and locality. But of course, saving both objective reality 
and locality cannot be without a price. It turns out that our
intermediate approach naturally
leads to a so-called {\em solipsistic}\footnote{The word ``solipsistic'' is borrowed
from philosophy, where it refers to the view that subjective mental experiences
are the only true reality.} 
reality, the meaning of which will become clear 
through the rest of the paper. 

To understand the basic idea, consider first a variant of the approach without objective
reality. This approach asserts that there is no reality except the observed
reality. Presumably, any observation ultimately happens in some part of a 
(conscious) brain, which is an object well localized in space. In this sense,
observations are local events. In particular, when an experimentalist 
(say, Alice) studies nonlocal EPR correlations, then all what she really observes are signals conveyed
to her brain, even if some of these signals originated from a distant apparatus
that measured spin of a distant member of the EPR pair. 
In this way Alice can insist that, from her point of view, 
entangled particles and the distant apparatus do not really exist.
From her point of view, all what exists are her observations, which are local.
For her this is the {\em only} reality. But since this is {\em her} reality,
it is not objective reality. In this way locality is saved with a price of loosing
objective reality.

Now our approach can be understood as a relatively small modification of the above.
What if the Alice's subjective observations are actually a result of some
{\em objective} physical processes in her brain? That would promote her subjective
reality into an objective one. And what if it is still true that other objects (such as
entangled particles and spatially separated measuring apparatuses
supposed to measure spins of these particles) are not real? 
That would retain locality. So in this way, it would be possible to have both objective reality
and locality. 

In this paper we construct an explicit quantitative model of objective reality inspired
by the qualitative idea above. Our model can be thought of as a variation of the
Bohmian hidden variable (HV) theory \cite{bohm,BMbook2,BMbook5}, with a difference that objective
existence and deterministic trajectories are ascribed only to those particles
which describe the degrees of freedom ultimately observed by the observer.
We show in detail how such a model is compatible with all measurable statistical predictions of
QM.

Before starting with a quantitative analysis, there is one additional important problem
to be addressed at a qualitative level. What if there is more than one observer,
say Alice and Bob? Alice could be egocentric by believing that only she really
exists, but Bob, who may be conscious of his own observations, would strongly disagree.
(Likewise, the author of this paper could believe that only he exists, 
but the reader of it would not buy it.) So, to avoid such an egocentric view of reality,
it is necessary to associate objective reality (in our model, particle trajectories)
with each conscious observer. Then each of them is local as an individual, 
but what if Alice observes one member of the EPR pair, while Bob, at the same time,
observes the other? To explain the EPR correlations, shouldn't real Alice and real Bob
mutually interact in a nonlocal way? 

While such a nonlocal interaction is one of the possibilities (which we 
shall demonstrate explicitly in a variant of our model), there is also a way to avoid it. 
The crucial observation is that Bob cannot determine experimentally
whether Alice is really conscious, and if she is, whether her state
of consciousness is consistent with his own. (And vice versa, of course.)
Therefore, the consistency with measurable EPR correlations does {\em not} require
their states of consciousness to be correlated. Consequently, HV's associated with Alice
can be independent of HV's associated with Bob, which avoids nonlocality.
Of course, Alice can hear that Bob tells her what his observations are. That information 
{\em about} Bob available to Alice can be correlated with other Alice's observations,
in agreement with predictions of QM. But the point is that an {\em Alice's} observation
is correlated with another {\em Alice's} observation, not with another 
Bob's observation. What Alice observes that Bob tells her that he observes 
is not necessarily what Bob really observes, and not because Bob is lying, but because
Bob as observed by Alice is not real. There is a real Bob, but this is not the one 
observed by Alice.


Such solipsistic reality is the price we pay for saving both reality and locality. 
One may think that the price is too big,
and we do not insist that it isn't. Yet, we do claim and insist that such a possibility is logically
consistent and compatible with all measured predictions of QM. 
In our opinion, this is a sufficient reason to explore such a possibility in more detail. 
Indeed, in the paper we shall see how such a view of reality naturally emerges from our simple 
quantitative model of HV's, fully compatible with measurable predictions of QM.

To avoid possible misunderstanding, 
when we say that solipsistic HV's are compatible with all measured predictions of QM,
we do not say that there are specific experiments which provide a direct evidence
for the existence of such HV's. Instead, we merely say that the measurable predictions of 
solipsistic HV's do not
seem to differ from those of other known interpretations, such as the orthodox no-HV interpretation 
or the Bohmian interpretation. Hence, in this paper we 
deal with the {\it general} theory of quantum measurements to understand
how the standard measurable predictions of QM can be reproduced from 
solipsistic HV's {\it in general}.
But we do not compare the predictions of solipsistic HV's with actual specific experiments,
because, with our present understanding of theory and experiments, 
we do not see how such specific experiments could help to determine whether the
solipsistic HV's are viable or not. Hopefully, it might change in the future.

The paper is organized as follows. In Sec.~\ref{SEC2} we review the essential and inessential
aspects of the Bohmian interpretation, in a manner that allows us to understand why exactly this
interpretation works and which aspects of it could potentially be abandoned.
Then in the central section, Sec.~\ref{SECsolip}, we explain how the particle ontology of the
Bohmian interpretation can be substantially reduced by replacing it with a new HV model, 
how that leads to a substantial reduction of nonlocality involved, and how such a model
can be given a natural solipsistic interpretation in terms of particle trajectories
describing the essential degrees of freedom of the observer. In Sec.~\ref{SEC4}
we generalize the model to the case of many observers. Finally, a qualitative conceptual
discussion of our results, as well as a conclusion, are given in Sec.~\ref{SEC5}.

\section{Essential and inessential aspects of Bohmian interpretation}
\label{SEC2}

The Bohmian interpretation is usually exposed by first presenting the equations 
of motion for particle trajectories, and then explaining why such trajectories
are compatible with all probabilistic predictions of QM \cite{bohm,BMbook2}. 
In our opinion, such a top-down approach
may not be the best way to teach Bohmian interpretation.
To understand more deeply why this interpretation works, 
it may be better to use a bottom-up approach in which
the ordering of exposition is reversed.

Such a bottom-up approach is what we do in this section.
We first present the interpretation-independent essentials of the quantum theory of measurements,
and then ask what kind of new objects should we have
and what kind of law should they satisfy in order to obtain
compatibility with measurable predictions resulting from QM. 
In this way it is easier to distinguish between essential and inessential aspect of
the Bohmian interpretation. This, of course, is valuable by itself, but our main motivation
for doing it is to prepare us for the next section
(Sec.~\ref{SECsolip}). Namely, with such an exposition of the Bohmian theory
it is easier to understand how the Bohmian theory can be replaced by a 
substantially different theory (sharing only the essential aspects with it),
which the subject of the next section is.

\subsection{Interpretation-independent essentials of the quantum theory of measurement}
\label{SEC2.1}

Suppose that one wants to measure the observable described by the Hermitian operator 
$\hat{K}$ with eigenstates $|k\rangle$ and eigenvalues $k$. 
To do that, one needs a measuring apparatus in an initial state
$|\Phi_0\rangle$ and interaction which causes a unitary transition of the form
\begin{equation}\label{e1}
 |k\rangle |\Phi_0\rangle \rightarrow |k'\rangle |\Phi_k\rangle ,
\end{equation}
where $|\Phi_0\rangle$ and $|\Phi_k\rangle$ are {\em macroscopically distinguishable}
states of the apparatus. In practice, macroscopic distinguishability means that 
$|\Phi_0\rangle$ and $|\Phi_k\rangle$ are many-particle states the wave functions of which
have a negligible overlap in the configuration space. To be more explicit, let 
${\bf x}_b$, $b=1,\ldots,n \gg 1$, denote positions of $n$ particles constituting the apparatus,
and let us introduce a shorthand notation 
\begin{equation}\label{e2}
({\bf x}_1,\ldots,{\bf x}_n) \equiv \vec{x} .
\end{equation} 
The negligible overlap of wave functions $\Phi_k(\vec{x}) \equiv \langle \vec{x}|\Phi_k \rangle$
means that
\begin{equation}\label{e3}
 \Phi_{k_1}(\vec{x}) \Phi_{k_2}(\vec{x}) \simeq 0 \;\;\;\; {\rm for} \;\;\;\;  k_1\neq k_2 .
\end{equation}
Thus, the unitary process (\ref{e1}) corresponds to a measurement of $\hat{K}$ in the following sense:
When the measured system is initially in the state $|k\rangle$, then the measuring apparatus
is finally found in the macroscopic state $|\Phi_k\rangle$. This is a reliable measurement owing to the fact
that the apparatus states are macroscopically distinguishable.
For later reference, we also note that the wave functions $\Phi_{k}(\vec{x})$ are normalized
\begin{equation}\label{normalization}
 \int d^{3n}x \, |\Phi_{k}(\vec{x})|^2 =1 ,
\end{equation}
where $d^{3n}x \equiv d^3x_1 \cdots d^3x_n$.

In the literature it is often assumed in (\ref{e1}) that $|k'\rangle=|k\rangle$, 
but such an assumption is neither always correct nor essential.
Measurements for which $|k'\rangle=|k\rangle$ are 
only a special case (sometimes referred to as measurements of the first kind), 
but in general $|k'\rangle \neq |k\rangle$.
For example, 
when $|k\rangle$ is a photon state, then its measurement usually destroys the photon
so that $|k'\rangle$ is the vacuum $|0\rangle$ for any initial $|k\rangle$.

An interesting question is why do we have a negligible overlap (\ref{e3}) in the 
position space, and not in some other space such as the momentum one?
A detailed answer is beyond the scope of the present paper (see, e.g., \cite{zeh-nopart}), 
but for our purposes it suffices to say that it can be explained by the theory of decoherence 
and the fact that interactions described by the Schr\"odinger equation are 
local in the position space, and not in some other space. This locality of interactions
in the Schr\"odinger equation is an interpretation-independent
fact of nature which, effectively, attributes a preferred 
role to the position basis in an interpretation-independent manner.
(This interpretation-independent locality should be distinguished from possible interpretation-dependent nonlocality of interactions between hidden variables, to be discussed later.)
But it should be stressed that the states $|k\rangle$ may correspond to any basis.
The position basis is preferred only for the macroscopic states of the measuring apparatus
(essentially because decoherence takes place only when a large number of degrees of freedom is 
involved \cite{decoh1,decoh2}), not for the microscopic states of the measured system.

The problem of measurement in QM becomes more challenging when the initial state
of the measured system is not an eigenstate $|k\rangle$, but a superposition
\begin{equation}\label{e4}
 |\psi\rangle = \sum_k c_k |k\rangle .
\end{equation}
In this case the initial state before the interaction with the measuring apparatus is 
$|\psi\rangle |\Phi_0\rangle$. The effect of interaction with the measuring apparatus
is determined completely by unitarity and Eq.~(\ref{e1}). Namely, unitarity and Eq.~(\ref{e1})
taken together give 
\begin{equation}\label{e5}
|\psi\rangle |\Phi_0\rangle \rightarrow \sum_k c_k |k'\rangle |\Phi_k\rangle .
\end{equation}

The Born rule applied to the right-hand side of (\ref{e5}) tells that the probability
for finding the detector in the state $ |\Phi_k\rangle$ is equal to $|c_k|^2$. 
This probability coincides with the Born-rule probability 
that the system will be found in the state
$|k\rangle$ in  (\ref{e4}). So given the Born rule, 
for most practical purposes we don't really need (\ref{e5}), i.e., we don't 
really need to care about the theory of quantum measurements.
For these practical purposes, (\ref{e4}) is enough.
Yet, the theory of quantum measurements outlined above 
is essential in any attempt to understand what really happens in a measurement.

Now the problem of measurement in QM can be reduced to the following question:
Given that the state after the measurement is the superposition on the right-hand side of
(\ref{e5}), why do we perceive that only one of the terms $c_k |k'\rangle |\Phi_k\rangle$
is physical? Or in more traditional language, why this superposition ``collapses'' to $|k'\rangle |\Phi_k\rangle$?
A part of the answer certainly lies in the fact that each of the terms contains a factor
$|\Phi_k\rangle$, because it makes all these terms macroscopically distinguishable.
This allows us to think of the right-hand side of (\ref{e5}) as a single macroscopic object
consisting of many distinguishable branches. Each branch evolves independently
and, for all practical purposes, behaves as if other branches did not exist. 
Thus, from the point of view of any particular branch, the other branches 
effectively do not exist. This sounds almost as an explanation of  
collapse (or more precisely, the {\em illusion} of collapse), but one should be careful.
We still need to answer one question: {\em Why should we take a view from the perspective of a 
branch as the physical one?}

Unfortunately, an interpretation-independent answer to that last
question does not seem to exist. To answer it we are forced to 
work within a paradigm of a particular
interpretation, which brings us to the next subsection. 

\subsection{The role of particle positions}
\label{SEC2.2}

To answer the last question of Sec.~\ref{SEC2.1}, one possibility
is to adopt a minimalistic approach, in which the wave function is postulated 
to be the {\em only} physical entity that exists. Thus, the branches 
discussed in Sec.~\ref{SEC2.1}, as parts of this physical entity, are physical
entities themselves. This possibility is better known under a more misleading name as
many-world interpretation \cite{mw1,mw2}. Yet, such a minimalistic approach
does not seem to be sufficient. At least, one needs some additional assumptions or axioms
in order to incorporate or explain the Born rule,
which may seem too {\it ad hoc} (see, e.g., \cite{redher} and references therein).

The Bohmian interpretation is a non-minimalistic approach to answer
the last question of Sec.~\ref{SEC2.1}. It usually starts from the axiom
that particles are physically real pointlike objects having specific deterministic trajectories, 
but here this will not be our starting point. Instead, we present a reversed
approach, in which we first try to find out what kind of objects do we {\em need},
and then construct such objects. 

The basic idea is that the point of view from 
a particular branch becomes physical because a particular
branch becomes {\em filled with} something physical. In other words, 
it is not the branch itself which is physical, but the entity which fills it.  
But what kind of entity could that be? It must be that an entity which fills
one branch does not fill any other branch (because otherwise we could not
say that only one branch is filled). Therefore, since the branches
are objects well localized in the configuration space, {\em the filling entity
must also be well localized in the configuration space}. In principle it could
be an object with a small but finite extension (for example, 
an experiment \cite{dehmelt} shows that the electron radius is smaller than $10^{-22}$ meters), 
but the simplest model is obtained if it is assumed to be a pointlike object of zero size.
This leads to the result that the physical entity is described by the position
\begin{equation}\label{e6}
 \vec{X} \equiv   ({\bf X}_1,\ldots,{\bf X}_n) 
\end{equation}
in the configuration space, 
where $n$ is the number of particles constituting the apparatus, as in (\ref{e2}).

This shows that (\ref{e6}) is a good candidate to be a real physical entity.  
However, it does not necessarily need to be the only real physical entity. 
As (\ref{e6}) is best visualized as positions in the 3-space of $n$ particles
constituting the apparatus, it seems natural to assume that the particles
constituting the apparatus are not fundamentally different from any other particles.
Thus, one may extend (\ref{e6}) by proposing that all particles have
positions 
\begin{equation}\label{e7}
 {\bf X}_1,\ldots,{\bf X}_n, {\bf X}_{n+1},\ldots,{\bf X}_N ,
\end{equation} 
where $N\gg n$ is the total number of particles in the Universe.

The above does not yet provide consistency with the statistical predictions of QM.
QM states that the probability density $\rho$ for particle positions 
${\bf x}_1,\ldots, {\bf x}_N$ at time $t$ is given by the
wave function of the Universe 
\begin{equation}
\Psi({\bf x}_1,\ldots, {\bf x}_N,t) =\langle {\bf x}_1,\ldots, {\bf x}_N|\Psi(t)\rangle  
\end{equation}
as
\begin{equation}\label{e8}
\rho({\bf x}_1,\ldots, {\bf x}_N,t) = |\Psi({\bf x}_1,\ldots, {\bf x}_N,t)|^2 .
\end{equation}
However, in practice one cannot experimentally test this equation directly, simply because
one cannot observe all particles in the Universe. What one really observes
is some macroscopic observable describing the measuring apparatus, which
we model with the positions (\ref{e6}). Therefore, a phenomenologically more interesting
probability density is the marginal (apparatus) probability density 
\begin{equation}\label{e9}
\rho^{\rm (appar)}({\bf x}_1,\ldots, {\bf x}_n,t) = 
\int d^3x_{n+1} \cdots d^3x_N \, \rho({\bf x}_1,\ldots,{\bf x}_n, {\bf x}_{n+1},\ldots,{\bf x}_N,t) ,
\end{equation}
obtained by averaging over the unobserved positions in (\ref{e8}).
For $t$ after the measurement, $|\Psi(t)\rangle$ is well modeled by a direct product of
the right-hand side of (\ref{e5}) with a state describing the rest of the Universe. Using this and
(\ref{e3}), Eq.~(\ref{e9}) gives
\begin{equation}\label{e11}
 \rho^{\rm (appar)}(\vec{x}) \simeq \sum_{k}
|c_k|^2  \, |\Phi_{k}(\vec{x})|^2
\end{equation}
for $t$ after the measurement. Thus, the probability to find the apparatus in the state $|\Phi_{k}\rangle$
is actually the probability to find the particle positions of the apparatus in the support of the wave function
$\Phi_{k}(\vec{x})$. (By support, we mean the region in the configuration space in which
$\Phi_{k}(\vec{x})$ is not negligible.)
From (\ref{e11}) and (\ref{normalization}) we see that this probability is
\begin{equation}\label{e12}
p_{k}=\int_{\rm supp \;\; \Phi_{k}} d^{3n}x \,  \rho^{\rm (appar)}(\vec{x}) \simeq |c_k|^2 ,
\end{equation}
in accordance with the Born rule.

Now we see what property a statistical ensemble of particle positions should have in order to be
compatible with probabilistic predictions of QM. The main requirement is that, at each time $t$,
the apparatus particles of the ensemble should have the distribution (\ref{e9}), because it is sufficient to reproduce (\ref{e12}). A simple way (but not necessarily the only way!) to achieve this is to require
that, at each $t$, {\em all} particles in the Universe should have the ensemble distribution (\ref{e8}). 
(Of course, there is only one Universe. In practice, a physical ensemble is realized by repeating many times
the ``same'' experiment, where ``same'' refers to the degrees of freedom which in practice
can be controlled by the experimentalist.)

\subsection{The role of particle trajectories}

In the last subsection we have identified some essential (and inessential)
elements needed to reproduce the Born rule (\ref{e12}). The most important result was
that we need particle positions with given ensemble distributions. 
But we are not done yet. The problem is that the distribution (\ref{e8}) is time dependent.
It implies that particle positions in any single member of the ensemble should be
time dependent as well. In other words, particles should have {\em trajectories}.

In principle, the trajectories could be stochastic. However, as (\ref{e8}) has a deterministic
time dependence (given by the Schr\"odinger equation), it is possible that 
the trajectories compatible with (\ref{e8})
could be deterministic as well. Furthermore, as (\ref{e8}) is a smooth
function of its arguments, one expects that particle trajectories are smooth too. 
In other words, particles are expected to have well defined velocities. But what that velocities
could be? To find an answer, assume that particle positions are distributed according to (\ref{e8})
at some initial time $t_0$ and ask what property should the velocities have in order
to retain the distribution (\ref{e8}) for any $t$?
Let ${\bf v}_a({\bf X}_1,\ldots, {\bf X}_N,t)$, $a=1,\ldots, N$, denotes the velocity
of the $a$'th particle at time $t$ when the positions of the particles at $t$ are
${\bf X}_1,\ldots, {\bf X}_N$. Then it is not difficult to see that the velocity function
${\bf v}_a({\bf x}_1,\ldots, {\bf x}_N,t)$ should satisfy the {\em continuity equation}
 \begin{equation}\label{e14}
 \frac{\partial \rho}{\partial t} + 
\sum_{a=1}^{N} \mbox{\boldmath $\nabla$}_a (\rho {\bf v}_a) =0 .
\end{equation}

For $\rho$ given by (\ref{e8}), the velocity function ${\bf v}_a$ 
satisfying (\ref{e14}) is not unique. Nevertheless,
one particularly simple choice is 
\begin{equation}
{\bf v}_a={\bf v}_a^{\rm (Bohm)} ,
\end{equation} 
where
\begin{equation}\label{e15}
 {\bf v}_a^{\rm (Bohm)} = \frac{-i\hbar}{2m_a} \,
\frac{\Psi^* \!\stackrel{\leftrightarrow\;}{ \mbox{\boldmath $\nabla$}_a }\!  \Psi}
       {\Psi^*\Psi} ,
\end{equation}
$m_a$ is the mass of the $a$'th particle,
and $f \!\stackrel{\leftrightarrow\;}{ \mbox{\boldmath $\nabla$}_a }\! h \equiv
f(\mbox{\boldmath $\nabla$}_a h) - (\mbox{\boldmath $\nabla$}_a f) h$.
Indeed, it is straightforward to show that (\ref{e15}) satisfies (\ref{e14}) when
$\Psi$ satisfies the Schr\"odinger equation in an arbitrary scalar potential
$U({\bf x}_1,\ldots, {\bf x}_N,t)$. The details of this derivation are not 
essential for our purposes, so we omit them \cite{bohm,BMbook2}.

Eq.~(\ref{e15}) can also be written in a more illuminating form in terms of the
velocity operator $\hat{\bf v}_a=\hat{\bf p}_a/m_a$, where 
$\hat{\bf p}_a=-i\hbar \mbox{\boldmath $\nabla$}_a$ is the momentum operator.
With this operator at hand, (\ref{e15}) can be written as
\begin{equation}\label{e24}
 {\bf v}_a^{\rm (Bohm)} = \frac{1}{2} \, \frac{ \Psi^*(\hat{\bf v}_a \Psi) + 
(\hat{\bf v}_a^{\dagger} \Psi^*)\Psi } {\Psi^*\Psi}
=\frac{ {\rm Re} (\Psi^*\hat{\bf v}_a \Psi)}{\Psi^*\Psi} ,
\end{equation}
where in the last equality we have used self-adjointness of the operator $\hat{\bf v}_a$.

There are many heuristic reasons to prefer (\ref{e15}) (equivalent to  (\ref{e24})) over many other choices
satisfying (\ref{e14}). The heuristic reasons include the analogy with classical Hamilton-Jacobi mechanics 
\cite{bohm,BMbook2}, Galilean invariance \cite{BMbook5}, and
the relation with weak measurement of velocity at a given position \cite{weak1,weak2}.  
Nevertheless, neither of these heuristic reasons is essential for reproducing the 
statistical predictions of QM.

Now we can see why the particle trajectories satisfy a nonlocal law.
The velocity of the $a$'th particle is
\begin{equation}\label{e16}
 \frac{d{\bf X}_a}{dt}={\bf v}_a({\bf X}_1,\ldots, {\bf X}_N,t) .
\end{equation}
The velocity of the particle at the position ${\bf X}_a$ at time $t$ may depend on the positions
of all other particles in the Universe at the same time, no matter how far they are from the
$a$'th particle. {\em That} is what is meant when said that the Bohmian interpretation
is a nonlocal HV theory. 
This nonlocality is a consequence of the fact that in (\ref{e7}) we have {\em assumed} that
all particles have positions, which is one of the essentials assumptions
of the Bohmian interpretation. Yet, as we have seen in (\ref{e6}) and (\ref{e12}), 
that assumption did not seem essential for the goal of reproducing the predictions of QM. 
The apparatus particle positions (\ref{e6}) could be sufficient, which could avoid nonlocality.
The aim of the rest of the paper is to explore that possible loophole in more detail.

\section{Reduced nonlocality from reduced particle ontology}
\label{SECsolip}

In Sec.~\ref{SEC2},
we have seen that only the apparatus particle positions (\ref{e6}), and not all
particle positions in the Universe (\ref{e7}), seemed essential for reproducing the
measurable predictions of QM. Could it be that only the apparatus particle positions are
real, i.e., that the particle ontology refers only to (\ref{e6}), not to (\ref{e7})?
In this section we explore the possibility of such a reduced particle ontology in more detail.

\subsection{General theory}
\label{SECsolip.1}

Let us start from the mathematical fact that (\ref{e15}) satisfies (\ref{e14}), 
but let us {\em not} assume particle trajectories (\ref{e16}).
We integrate (\ref{e14}) over $\int d^3x_{n+1} \cdots d^3x_N$, which leads to
\begin{eqnarray}\label{e17}
& \displaystyle\frac{ \partial \rho^{\rm (appar)}({\bf x}_1,\ldots, {\bf x}_n,t) }{\partial t} +
\sum_{b=1}^{n} \mbox{\boldmath $\nabla$}_b
\int d^3x_{n+1} \cdots d^3x_N \, \rho {\bf v}_b^{\rm (Bohm)}   &
\nonumber \\
& + \displaystyle\sum_{a=n+1}^{N}  \int d^3x_{n+1} \cdots d^3x_N \,
\mbox{\boldmath $\nabla$}_a (\rho {\bf v}_a^{\rm (Bohm)})  =0 , &
\end{eqnarray}
where $\rho^{\rm (appar)}$ is defined by (\ref{e9}).
By the Gauss theorem, the last term in (\ref{e17}) reduces to a surface integral, 
which vanishes. Therefore, (\ref{e17}) can be written as
\begin{equation}\label{e18}
 \frac{ \partial \rho^{\rm (appar)} (\vec{x},t) }{\partial t} +
\sum_{b=1}^{n} \mbox{\boldmath $\nabla$}_b
[\rho^{\rm (appar)}(\vec{x},t)  {\bf v}_b^{\rm (appar)} (\vec{x},t) ] =0 ,
\end{equation}
where 
\begin{equation}\label{e19}
 {\bf v}_b^{\rm (appar)} (\vec{x},t) \equiv 
\displaystyle\frac{ \displaystyle\int d^3x_{n+1} \cdots d^3x_N \, 
\rho(\vec{x},{\bf x}_{n+1},\ldots,{\bf x}_N,t)
{\bf v}_b^{\rm (Bohm)}(\vec{x},{\bf x}_{n+1},\ldots,{\bf x}_N,t) }
{ \rho^{\rm (appar)}(\vec{x},t) } .
\end{equation}
Eq.~(\ref{e18}) has the form of a continuity equation for the apparatus probability density
$\rho^{\rm (appar)}$. Therefore, the quantum-mechanical probability (\ref{e12})
of finding the apparatus in the state $|\Phi_{k}\rangle$ can be reproduced 
by postulating that the apparatus particles have trajectories
\begin{equation}\label{e20}
 \frac{d{\bf X}_b}{dt}={\bf v}_b^{\rm (appar)}({\bf X}_1,\ldots, {\bf X}_n,t) ,
\end{equation}
where $b=1,\ldots, n$.

We stress that the trajectories (\ref{e20}) are very different from the Bohmian trajectories
(\ref{e16}). First, {\em only the apparatus particles have trajectories} in (\ref{e20}).
Second, (\ref{e20}) is much less nonlocal than (\ref{e16}), in the sense that 
{\em the velocity of the $b$'th particle of the apparatus in (\ref{e20}) depends only on the other particle
positions of the apparatus, not on the positions of any other particles in the Universe}.

For the sake of deeper understanding of Eqs.~(\ref{e9}) and (\ref{e19}), it is also useful
to write them in a different form, in terms of density matrices and partial traces.
The density matrix associated with the state $|\Psi(t)\rangle$ is the operator
\begin{equation}\label{e21}
 \hat{\rho}(t)=|\Psi(t)\rangle \langle \Psi(t)| .
\end{equation}
Its matrix elements in the configuration-space basis are 
 \begin{eqnarray}\label{e22}
 \rho({\bf x}_1,\ldots, {\bf x}_N ; {\bf x}'_1,\ldots, {\bf x}'_N ; t) & = & 
\langle {\bf x}_1,\ldots, {\bf x}_N |\Psi(t)\rangle \langle \Psi(t)| {\bf x}'_1,\ldots, {\bf x}'_N \rangle
\nonumber \\
& = & 
\Psi({\bf x}_1,\ldots, {\bf x}_N ,t)  \Psi^*({\bf x}'_1,\ldots, {\bf x}'_N ,t) .
 \end{eqnarray}
Thus, (\ref{e8}) is nothing but the diagonal matrix elements of (\ref{e21}).
Similarly, (\ref{e9}) is nothing but the diagonal matrix elements of
the {\em reduced density matrix}
\begin{equation}\label{e23}
 \rho^{\rm (appar)}(\vec{x};\vec{x}';t)=[{\rm Tr}_{\mbox{\scriptsize (no-appar)}} 
\hat{\rho}](\vec{x};\vec{x}';t) ,
\end{equation}
where ${\rm Tr}_{\mbox{\scriptsize (no-appar)}}$ denotes the partial trace
over all no-apparatus degrees of freedom. 
The numerator of (\ref{e19}) can be written in a similar form, by using (\ref{e24})
written as $\rho {\bf v}_b^{\rm (Bohm)}={\rm Re} (\Psi^*\hat{\bf v}_b \Psi)$.
In this way one finally obtains that (\ref{e19}) can be written as
\begin{equation}\label{e25}
{\bf v}_b^{\rm (appar)} (\vec{x},t) = 
\frac{ {\rm Re} [ {\rm Tr}_{\mbox{\scriptsize (no-appar)}} 
\hat{{\bf v}}_b \hat{\rho} ] (\vec{x};\vec{x};t) } 
{[{\rm Tr}_{\mbox{\scriptsize (no-appar)}} \hat{\rho} ] (\vec{x};\vec{x};t) } .
\end{equation}

The form (\ref{e25}) is a very convenient one because the properties
of partial traces are well known from quantum-information theory.
In particular, 
matrix elements of operators of the form appearing in (\ref{e25}) obey locality in the following sense:
The dynamics of $|\Psi(t)\rangle$ is governed by a Hamiltonian $\hat{H}$, 
i.e., $\hat{H}$ influences the $\Psi$-degrees of freedom.
If the no-apparatus $\Psi$-degrees of freedom are influenced by a local $\hat{H}$
that does not influence the apparatus $\Psi$-degrees of freedom, then the partial traces
of the form appearing in (\ref{e25}) do not depend on these influences.
For the sake of completeness, we derive this well-known fact in the Appendix.
In quantum-information theory, this fact is used as a proof that nonlocal EPR correlations cannot be used
for a nonlocal transmission of information. Likewise, for our purpose it proves that 
the particle velocities (\ref{e20}) do not depend on such influences on the no-apparatus $\Psi$-degrees
of freedom.

Of course, a macroscopic apparatus has a finite spatial extension and contains $n>1$ particle trajectories,
so it is not perfectly local. Yet, compared with nonlocality in Bohmian mechanics,
the present model has a substantially reduced nonlocality. First, the spatial extension
of a typical apparatus is much smaller than the spatial distance between entangled particles
(see Sec.~\ref{SECsolip.2} for an example). Second, decoherence involved in a typical
macroscopic apparatus destroys entanglement between most pieces of the apparatus,
so that effective entanglement is present only at a microscopic level. For instance, 
electrons within the same molecule are typically entangled, but usually there is no entanglement 
between different molecules. In this sense, nonlocality is typically reduced down to microscopic distances inside the apparatus.

\subsection{Example: Measurement of EPR correlations}
\label{SECsolip.2}

\begin{figure}[b]
\centerline{\includegraphics[width=14cm]{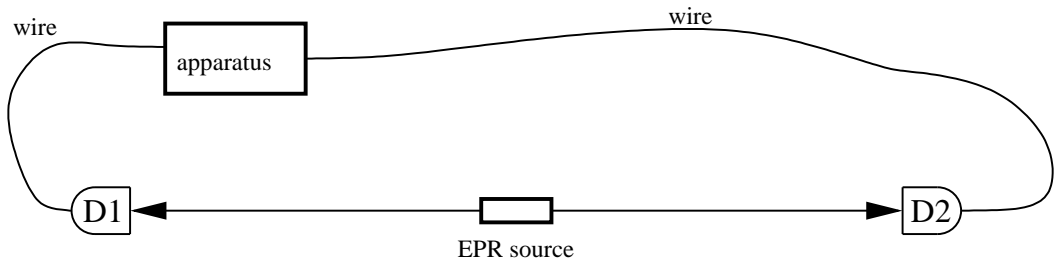}}
\caption{\label{epr-fig}Sketch of a setup for measurement of EPR correlations. 
The entangled EPR pair is produced by the
EPR source. Individual particles are detected by the detectors D1 and D2. Information about 
the individual detection results are transmitted by wires to the measuring apparatus which 
measures the correlations between the individual detection results.}
\end{figure}

As an example, consider a measurement of EPR correlations by a setup 
sketched at Fig.~\ref{epr-fig}. 
Before the detections at detectors D1 and D2, the entangled EPR state is
\begin{equation}\label{e4.1}
 |\psi\rangle = \frac{|\!\uparrow\rangle  |\!\downarrow\rangle + |\!\downarrow\rangle  |\!\uparrow\rangle}
{\sqrt{2}}
= \sum_{k_1,k_2} c_{k_1 k_2} |k_1\rangle |k_2\rangle ,
\end{equation}
where $|k_1\rangle$ and $|k_2\rangle$ are eigenstates of the observables to be detected by detectors D1 and D2, respectively.
After the detections at D1 and D2 and measurement by the apparatus, the total state can be modeled
by a state of the form
\begin{equation}\label{e5.1}
|\Psi\rangle = \sum_ {k_1,k_2} c_{k_1 k_2}|k_1\rangle |k_2\rangle  |D1_{k_1}\rangle |D2_{k_2}\rangle
|\Phi_{k_1,k_2}\rangle ,
\end{equation} 
where $|D1_{k_1}\rangle$, $|D2_{k_2}\rangle$, and $|\Phi_{k_1,k_2}\rangle$
are macroscopically distinguishable states of the detector D1, detector D2, and measuring
apparatus, respectively. (For simplicity, we have suppressed the factors corresponding
to the states of wires and the EPR source.)

Now it is not difficult to see how the hidden variables (particle trajectories) of Sec.~\ref{SECsolip.1}
avoid nonlocality of the Bell theorem \cite{bell}.
The Bell theorem assumes that there are some hidden variables (HV's) associated with the
entangled measured system (described in the basis $|k_1\rangle |k_2\rangle$) and/or
separated detectors (described in the basis $|D1_{k_1}\rangle |D2_{k_2}\rangle$).
If such HV's exist, then the Bell theorem asserts that they must be nonlocal.
However, such HV's do {\em not} exist in our approach. Instead, HV's are attributed
{\em only} to the local measuring apparatus, described in the basis $|\Phi_{k_1,k_2}\rangle$.

\subsection{Solipsistic interpretation}

To save locality associated with the measurement of EPR correlations in Sec.~\ref{SECsolip.2},
we have associated HV's (particle trajectories) only with the macroscopic apparatus that measures
the correlations, not with the macroscopic particle detectors. This corresponds to the
view that the macroscopic apparatus measuring correlations is objectively real (ontological), while
the particle detectors are not. This seems very odd for two reasons. First, 
from the point of view of our everyday experience, both seem equally real. Second, if some 
but not all macroscopic objects are real, then how to know in general which ones are real and which ones
are not? 

To resolve this conceptual puzzle we would need some macroscopic apparatus which, in some sense,
could be believed to be ``more real'' than others. Fortunately, there {\em is} 
a candidate for such an apparatus -- 
the brain of the observer. All someone's knowledge about the world is ultimately represented by a state
of his/her brain. In this sense, if any material object could potentially be believed to be more real
than others, then it is the brain. Or more precisely, not necessarily all parts of the brain,
but only those essential parts which are ultimately responsible for consciousness. Neurosciences
cannot yet unequivocally identify what that essential neural correlates of consciousness are, but
the research in that direction is ongoing \cite{NCC}.

This leads to the solipsistic interpretation, according to which the generic measuring ``apparatus'' 
discussed so far is actually -- the conscious observer. The particle trajectories are associated only
with the parts of brain which are essential for the creation of consciousness. 

Of course, such a proposal is certainly not without problems. Perhaps it even creates more 
problems than solves. For that reason, our model should not be taken
too seriously as a candidate for the final theory of physical reality.
Yet, at least it should be taken as an explicit counterexample to the view that
local reality is not compatible with QM. Our model explicitly shows that local reality is
possible at least in principle, even if the model is viewed as a toy model only.

One particularly difficult question with the model is the following.
Presumably, a brain becomes conscious at some particular early time of its evolution.
Therefore, our model would require that the brain particles get their objective positions
at that particular time. But how exactly would that happen? Our model says
nothing about that. An enriched model which would describe that phenomenon
is certainly conceivable, but nothing simple or natural of that form comes to our mind. 
Yet, we stress that this problem is independent of the fact that our model avoids nonlocality
related to measurements of EPR correlations. 

Thus, we conclude that local solipsistic reality compatible with measurable predictions of QM 
is at least logically possible, even if not fully satisfying in the present form. However
we are note done yet, because our analysis so far refers 
to a single observer only. We still need to generalize it to the case of many observers, 
which we do in the next section.

\section{Generalization to many observers}
\label{SEC4}

As discussed qualitatively in Introduction, it does not seem reasonable to associate particle trajectories
with only one observer. Instead, particle trajectories should be associated with all of them. 
Therefore, in this section we generalize the results obtained so far to the case of many observers.
For simplicity, we consider the case of two observers (Alice and Bob), but present it in such a form that the generalization to an arbitrary number of observers is obvious.

As an example, consider an entangled pair of particles in the state
\begin{equation}\label{e4.2}
 |\psi\rangle = \frac{ |\!\uparrow_1\rangle \otimes |\!\downarrow_2\rangle + 
|\!\downarrow_1\rangle  \otimes |\!\uparrow_2\rangle }{\sqrt{2}} ,
\end{equation} 
where the labels $1$ and $2$ denote particle states localized in vicinity of the observers Alice and Bob, respectively. 
(Mathematically, $|a\rangle |b\rangle$ is the same as $|a\rangle \otimes |b\rangle$, but we 
use the symbol $\otimes$ when it helps equation look more intelligible.)
Let Alice perform a measurement which determines whether the state in her vicinity is
$|\!\uparrow_1\rangle$ or  $|\!\downarrow_1\rangle$. Likewise, let Bob
perform a measurement which determines whether the state in his vicinity is
$|\!\uparrow_2\rangle$ or  $|\!\downarrow_2\rangle$. Then the total state after the measurement can be modeled as
\begin{equation}\label{e30}
 |\Psi\rangle =  \frac{1}{\sqrt{2}}\, |\!\uparrow_1\rangle |\Phi^{\rm (A)}_{\uparrow}\rangle \otimes 
 |\!\downarrow_2\rangle |\Phi^{\rm (B)}_{\downarrow}\rangle +
\frac{1}{\sqrt{2}}\, |\!\downarrow_1\rangle |\Phi^{\rm (A)}_{\downarrow}\rangle \otimes 
 |\!\uparrow_2\rangle |\Phi^{\rm (B)}_{\uparrow}\rangle ,
\end{equation}
where $|\Phi^{\rm (A)}_{\uparrow}\rangle$ corresponds to the case that Alice observes spin up, 
$|\Phi^{\rm (B)}_{\downarrow}\rangle$ corresponds to the case that Bob observes spin down, etc.

Now we want to associate $n_{\rm A}$ particle trajectories with the observer Alice and 
$n_{\rm B}$ particle trajectories with the observer Bob, with velocities
\begin{equation}\label{e31}
\frac{d{\bf X}_b}{dt}={\bf v}_b , \;\;\;\; b=1,\ldots, n_{\rm A}+n_{\rm B} .
\end{equation}
The problem reduces to finding the appropriate quantities ${\bf v}_b$
as functions of particle positions ${\bf x}_1, \ldots,{\bf x}_{n_{\rm A}+n_{\rm B}}$. 
We shall see that there are two
different approaches to choose these functions: a nonlocal approach and a local one. 

\subsection{Nonlocal approach}

An obvious possibility is to introduce a notation that generalizes (\ref{e6}) as 
\begin{equation}\label{e6.1}
 \vec{X} \equiv   ({\bf X}_1,\ldots,{\bf X}_{n_{\rm A}+n_{\rm B}}) ,
\end{equation}
and to generalize  (\ref{e25}) to 
\begin{equation}\label{e25.1}
{\bf v}_b^{\rm (AB)} (\vec{x},t) = 
\frac{ {\rm Re} [ {\rm Tr}_{\mbox{\scriptsize (no-AB)}} 
\hat{{\bf v}}_b \hat{\rho} ] (\vec{x};\vec{x};t) } 
{[{\rm Tr}_{\mbox{\scriptsize (no-AB)}} \hat{\rho} ] (\vec{x};\vec{x};t) } ,
\end{equation}
where $b=1,\ldots, n_{\rm A}+n_{\rm B}$ and ${\rm Tr}_{\mbox{\scriptsize (no-AB)}}$
denotes the partial trace over all degrees of freedom except those corresponding to
particle positions associated with the observers Alice and Bob.
This implies that the joint probability density for Alice's and Bob's particle positions
\begin{equation}\label{e34}
 \rho^{\rm (AB)} (\vec{x},t) = [{\rm Tr}_{\mbox{\scriptsize (no-AB)}} \hat{\rho} ] (\vec{x};\vec{x};t)
\end{equation}
satisfies the joint continuity equation
\begin{equation}\label{e18.1}
 \frac{ \partial \rho^{\rm (AB)} (\vec{x},t) }{\partial t} +
\sum_{b=1}^{n_{\rm A}+n_{\rm B}} \mbox{\boldmath $\nabla$}_b
[\rho^{\rm (AB)}(\vec{x},t)  {\bf v}_b^{\rm (AB)} (\vec{x},t) ] =0 .
\end{equation}
Thus it is consistent to propose that particles have velocities (\ref{e31}) with
\begin{equation}
 {\bf v}_b={\bf v}_b^{\rm (AB)} .
\end{equation}

In particular, (\ref{e34}) assigns a zero probability that, at the same time, 
Alice's particles have positions in the first branch 
$|\!\uparrow_1\rangle |\Phi^{\rm (A)}_{\uparrow}\rangle \otimes 
 |\!\downarrow_2\rangle |\Phi^{\rm (B)}_{\downarrow}\rangle$ of (\ref{e30}),
while Bob's particles have positions in its the second branch
$|\!\downarrow_1\rangle |\Phi^{\rm (A)}_{\downarrow}\rangle \otimes 
 |\!\uparrow_2\rangle |\Phi^{\rm (B)}_{\uparrow}\rangle$.
Instead, either both Alice's and Bob's particles are in the first branch,
or both Alice's and Bob's particles are in the second branch.
Only one branch contains real particles. This means that the motion
of Alice's particles is correlated with the motion of Bob's particles.
Indeed, this is reflected in (\ref{e25.1}) with (\ref{e6.1}), which shows
that a velocity of an Alice's particle may depend not only on other Alice's particle positions,
but also on Bob's particle positions. Therefore, the obvious approach studied
in this subsection is nonlocal, which is not what we want. 
For that reason, in the next subsection we consider a different approach.

\subsection{Local approach}

Now instead of (\ref{e6.1}) we introduce a different notation
\begin{equation}\label{e6.2}
 \vec{X}^{\rm (A)} \equiv   ({\bf X}^{\rm (A)}_1,\ldots,{\bf X}^{\rm (A)}_{n_{\rm A}}) , \;\;\;\;\;\;
\vec{X}^{\rm (B)} \equiv   ({\bf X}^{\rm (B)}_1,\ldots,{\bf X}^{\rm (B)}_{n_{\rm B}}) .
\end{equation}
Similarly, instead of (\ref{e25.1}) we introduce two independent velocities
\begin{equation}\label{e25.2a}
{\bf v}_b^{\rm (A)} (\vec{x}^{\rm (A)},t) = 
\frac{ {\rm Re} [ {\rm Tr}_{\mbox{\scriptsize (no-A)}} 
\hat{{\bf v}}_b \hat{\rho} ] (\vec{x}^{\rm (A)};\vec{x}^{\rm (A)};t) } 
{[{\rm Tr}_{\mbox{\scriptsize (no-A)}} \hat{\rho} ] (\vec{x}^{\rm (A)};\vec{x}^{\rm (A)};t) } ,
\end{equation}
for $b=1,\ldots, n_{\rm A}$, 
and 
\begin{equation}\label{e25.2b}
{\bf v}_b^{\rm (B)} (\vec{x}^{\rm (B)},t) = 
\frac{ {\rm Re} [ {\rm Tr}_{\mbox{\scriptsize (no-B)}} 
\hat{{\bf v}}_b \hat{\rho} ] (\vec{x}^{\rm (B)};\vec{x}^{\rm (B)};t) } 
{[{\rm Tr}_{\mbox{\scriptsize (no-B)}} \hat{\rho} ] (\vec{x}^{\rm (B)};\vec{x}^{\rm (B)};t) } ,
\end{equation}
for $b=1,\ldots, n_{\rm B}$. The corresponding probability densities
\begin{equation}\label{density-a}
 \rho^{\rm (A)} (\vec{x}^{\rm (A)},t) = 
[{\rm Tr}_{\mbox{\scriptsize (no-A)}} \hat{\rho} ] (\vec{x}^{\rm (A)};\vec{x}^{\rm (A)};t) ,
\end{equation}
\begin{equation}\label{density-b}
 \rho^{\rm (B)} (\vec{x}^{\rm (B)},t) = 
[{\rm Tr}_{\mbox{\scriptsize (no-B)}} \hat{\rho} ] (\vec{x}^{\rm (B)};\vec{x}^{\rm (B)};t) ,
\end{equation}
satisfy two independent continuity equations
\begin{equation}\label{e18.2a}
 \frac{ \partial \rho^{\rm (A)} (\vec{x}^{\rm (A)},t) }{\partial t} +
\sum_{b=1}^{n_{\rm A}} \mbox{\boldmath $\nabla$}_b
[\rho^{\rm (A)}(\vec{x}^{\rm (A)},t)  {\bf v}_b^{\rm (A)} (\vec{x}^{\rm (A)},t) ] =0 ,
\end{equation}
\begin{equation}\label{e18.2b}
 \frac{ \partial \rho^{\rm (B)} (\vec{x}^{\rm (B)},t) }{\partial t} +
\sum_{b=1}^{n_{\rm B}} \mbox{\boldmath $\nabla$}_b
[\rho^{\rm (B)}(\vec{x}^{\rm (B)},t)  {\bf v}_b^{\rm (B)} (\vec{x}^{\rm (B)},t) ] =0 .
\end{equation}
Therefore the Alice's particle velocities 
\begin{equation}\label{e31a}
\frac{d{\bf X}^{\rm (A)}_b}{dt}={\bf v}^{\rm (A)}_b (\vec{X}^{\rm (A)},t) , \;\;\;\; 
b=1,\ldots, n_{\rm A} ,
\end{equation}
are compatible with the density (\ref{density-a}), while the Bob's particle velocities
\begin{equation}\label{e31b}
\frac{d{\bf X}^{\rm (B)}_b}{dt}={\bf v}^{\rm (B)}_b (\vec{X}^{\rm (B)},t) , \;\;\;\; 
b=1,\ldots, n_{\rm B} ,
\end{equation}
are compatible with the density (\ref{density-b}). Clearly, the particle trajectories
(\ref{e31a}) and (\ref{e31b}) save locality, in the sense that the velocities of Alice's
particles do not depend on positions of Bob's particles, and vice versa. 

There is, however, a price to be payed for saving locality. Now the motion of
the Alice's particles is not correlated with the motion of the Bob's particles.
The particle trajectories are no longer compatible with the joint probability density
(\ref{e34}), but only with the separate probability densities (\ref{density-a}) and
(\ref{density-b}).
Consequently, the Alice's particles do not necessarily need to be in the same branch
as Bob's particles. 

The last feature is particularly interesting when Alice and Bob measure the same observable.
For instance, let the measured system be described by the state
\begin{equation}\label{e4.3}
 |\psi\rangle = \frac{ |\!\uparrow\rangle  +  |\!\downarrow\rangle }{\sqrt{2}} ,
\end{equation}
and let both Alice and Bob measure whether the system is in the state $|\!\uparrow\rangle$ 
or $|\!\downarrow\rangle$. Then the total state after the measurement has the form
\begin{equation}\label{e30.2}
 |\Psi\rangle =  \frac{1}{\sqrt{2}}\, |\!\uparrow\rangle |\Phi^{\rm (A)}_{\uparrow}\rangle  
 |\Phi^{\rm (B)}_{\uparrow}\rangle +
\frac{1}{\sqrt{2}}\, |\!\downarrow\rangle |\Phi^{\rm (A)}_{\downarrow}\rangle 
 |\Phi^{\rm (B)}_{\downarrow}\rangle .
\end{equation}
It may happen that the Alice's particles are in the first branch (i.e., in the 
support of $\Phi^{\rm (A)}_{\uparrow}(\vec{x}^{\rm (A)}$), while
the Bob's particles are in the second branch (i.e., in the 
support of $\Phi^{\rm (B)}_{\downarrow}(\vec{x}^{\rm (B)}$).
This means that Alice and Bob may not agree on whether the measured system
is in the state $|\!\uparrow\rangle$ or $|\!\downarrow\rangle$.
This may seem to be in contradiction with the measurable predictions of QM,
but it is not. Neither Alice nor Bob can observe any contradiction, simply because
Bob's particle positions are {\em hidden} variables for Alice, just as
Alice's particle positions are {\em hidden} variables for Bob.
This, indeed, is fully compatible with our solipsistic interpretation
of the trajectories: Bob's mental experiences 
are hidden to Alice, just as Alice's mental experiences are hidden to Bob.

Finally note that our result that different observers may live in different branches of the wave function
is very similar to the many-world interpretation \cite{mw1,mw2}, briefly discussed
in Sec.~\ref{SEC2.2}. Yet, there is one crucial difference. In the many-world interpretation, 
there is a copy of each observer in any of the branches. In our solipsistic interpretation,
for each observer there is only one copy living in only one of the branches.

\section{Discussion and conclusion}
\label{SEC5}

Bohmian mechanics reproduces measurable predictions of QM by associating 
particle trajectories with all coordinates in the configuration space on which
the wave function of the Universe depends. 
In this paper we have have seen that measurable predictions of QM can also be reproduced
by associating particle trajectories not with all these coordinates, but only those
which describe the essential degrees of freedom of the observers. 
Such a restricted particle ontology substantially reduces nonlocality involved
in the equations of motion for particle trajectories. Namely, nonlocal influences
between particles are restricted to distances corresponding to the size of the observer. Moreover,
quantum coherence in a brain cannot sustain at macroscopic distances \cite{tegmark},
which can be used to argue that nonlocality is effectively reduced down to microscopic distances 
within the observer. In this sense, such a solipsistic theory of hidden variables
can be considered local when compared with Bohmian mechanics.

There is, however, one important conceptual question to be addressed.
Even though the theory is used to calculate the particle trajectories, the theory
still uses the wave function of the Universe in the full configuration space.
This wave function is not a separable entity. So does it mean that
the theory is still nonlocal?

The answer is rather subtle and depends on definition of ``nonlocality''. 
First, even though nonlocality and nonseparability are closely
related concepts, they are not exactly the same. While the notion of nonlocality
involves an action at a distance (which is very manifest in Bohmian mechanics),
the notion of nonseparability of the wave function does not involve
action at a distance. Second, the notions of locality and nonlocality both refer to
a property {\em in} the 3-space. (Indeed, to discuss whether local or nonlocal
reality exists one must first assume that at least the 3-space exists, because
otherwise such a discussion does not make sense \cite{norsen}.)
On the other hand, the wave function, living in the configuration space,
is not even defined in the 3-space. In this sense, the wave function is neither
local nor nonlocal. 

Third, and most important, even if the notion of ``nonlocality'' is redefined such that
nonseparability of the wave function is also viewed as a kind of nonlocality, 
there is a subtle question whether
the wave function is {\em ontological} in our solipsistic HV model. 
It is certainly ``less ontological'' than the wave function in the
many-world interpretation (MWI), in the sense that the wave function is the main
ontological entity in MWI, while in our solipsistic HV model the wave function
only has an auxiliary role, as a quantity that serves to calculate the main
ontological entities (particle trajectories) which, in turn, obey local laws. 
Thus, as far as the wave function is
``less ontological'' in the solipsistic model than in MWI, one can say that
the solipsistic model is also ``less nonlocal'' than MWI.

Of course, one could also argue that the mere fact that one needs the 
nonseparable wave function
to calculate something is a sufficient reason to conclude that the theory is ``nonlocal''.
But then {\em any} interpretation of QM is ``nonlocal'' in that sense, including the interpretations
\cite{mermcor,zeil,rov2} that deny the existence of objective reality.
And even then our solipsistic HV model has a value as a bridge interpolating between two 
confronting views of QM, namely those with \cite{bell,durretal,norsen} and without \cite{mermcor,zeil,rov2}
objective reality.

To conclude, the solipsistic HV interpretation of QM 
offers a novel view of QM with some advantages and disadvantages with respect to other 
existing interpretations. We believe that it significantly contributes to the
general conceptual understanding of QM, even if it does not accurately describe what
is ``really'' going on behind the abstract laws of QM.

\section*{Acknowledgements}


This work was supported by the Ministry of Science of the
Republic of Croatia under Contract No.~098-0982930-2864.

\appendix

\section{Partial traces for non-interacting subsystems}

Let $A$ and $B$ be two quantum subsystems with bases in the corresponding Hilbert spaces
$\{ |k_A\rangle \}$ and  $\{ |k_B\rangle \}$, respectively. The basis for the total Hilbert space
is the set $\{ |k_A\rangle \otimes |k_B\rangle \}$. Let as assume that there is no mutual
interaction between the two subsystems. Then the total (possibly time-dependent) Hamiltonian
has the form
\begin{equation}
 \hat{H}(t)=\hat{H}_A(t)+\hat{H}_B(t) ,
\end{equation}
where $\hat{H}_A(t)$ and $\hat{H}_B(t)$ act only on the systems $A$ and $B$, respectively:
\begin{equation}
 \hat{H}(t) [ |k_A\rangle \otimes |k_B\rangle ]= \hat{H}_A(t) |k_A\rangle \otimes |k_B\rangle +
|k_A\rangle \otimes \hat{H}_B(t) |k_B\rangle .
\end{equation}
Hence, the unitary time-evolution operator
\begin{equation}
 \hat{U}(t)=T e^{-\frac{i}{\hbar} \int_{t_0}^t dt' \hat{H}(t') } ,
\end{equation}
where $T$ denotes the time ordering, factorizes as
\begin{equation}
 \hat{U}(t)=\hat{U}_A(t) \otimes \hat{U}_B(t) ,
\end{equation}
where
\begin{equation}
 \hat{U}_A(t)=T e^{-\frac{i}{\hbar} \int_{t_0}^t dt' \hat{H}_A(t') } , \;\;\;\;
\hat{U}_B(t)=T e^{-\frac{i}{\hbar} \int_{t_0}^t dt' \hat{H}_B(t') } ,
\end{equation}
are unitary time-evolution operators for separate subsystems.
Hence, an arbitrary (pure) time-dependent state of the total system
\begin{equation}
 |\Psi(t)\rangle = \hat{U}(t)  |\Psi(t_0)\rangle
\end{equation}
can be written in the form
\begin{equation}\label{app-psi}
 |\Psi(t)\rangle = \sum_{k_A,k_B} c_{k_A k_B} \hat{U}_A(t) |k_A\rangle \otimes
\hat{U}_B(t) |k_B\rangle . 
\end{equation}

Now let
\begin{equation}
 \hat{\rho}(t)=|\Psi(t)\rangle \langle \Psi(t)|
\end{equation}
be the density matrix associated with (\ref{app-psi}) and let $\hat{O}_A$ be some 
operator acting on system $A$.
Their product is
\begin{equation}
 \hat{O}_A \hat{\rho}(t)= \sum_{k_A,k_B} \sum_{k'_A,k'_B} c_{k_A k_B} c^*_{k'_A k'_B}
\hat{O}_A \hat{U}_A(t) |k_A\rangle \langle k'_A |\hat{U}^{\dagger}_A(t) \otimes
\hat{U}_B(t) |k_B\rangle \langle k'_B |\hat{U}^{\dagger}_B(t) .
\end{equation}
The partial trace of it
\begin{equation}\label{app-tr0}
 {\rm Tr}_B \hat{O}_A \hat{\rho}(t)=\sum_{k''_B} \langle k''_B | \hat{O}_A \hat{\rho}(t) | k''_B\rangle
\end{equation}
is then given by
\begin{equation}\label{app-tr}
{\rm Tr}_B \hat{O}_A \hat{\rho}(t) = \sum_{k_A,k_B} \sum_{k'_A,k'_B} c_{k_A k_B} c^*_{k'_A k'_B}
\hat{O}_A \hat{U}_A(t) |k_A\rangle \langle k'_A |\hat{U}^{\dagger}_A(t)
\, \Delta_{k_B k'_B}(t) ,
\end{equation}
where
\begin{eqnarray}
 \Delta_{k_B k'_B}(t) & = & \sum_{k''_B} 
\langle k''_B |  \hat{U}_B(t) |k_B\rangle \langle k'_B |\hat{U}^{\dagger}_B(t)  | k''_B\rangle
\nonumber \\
& = & \sum_{k''_B} 
\langle k'_B |\hat{U}^{\dagger}_B(t)  | k''_B\rangle \langle k''_B |  \hat{U}_B(t) |k_B\rangle
\nonumber \\
& = & \langle k'_B |\hat{U}^{\dagger}_B(t) \hat{U}_B(t) |k_B\rangle
\nonumber \\
& = & \langle k'_B |k_B\rangle = \delta_{k_B k'_B} .
\end{eqnarray}
This shows that $\Delta_{k_B k'_B}$ appearing in (\ref{app-tr}) depends neither on time
nor on $\hat{U}_B(t)$. Consequently, the partial trace (\ref{app-tr})
does not depend on $\hat{H}_B(t)$. Physically, it means that the partial trace
(\ref{app-tr0}) does not depend on physical influences on the subsystem $B$.

\end{document}